\documentclass[aps,amsfonts,prl,twocolumn,showpacs]{revtex4-1}

\usepackage{subfigure}
\usepackage{wrapfig}
\usepackage{textcomp}
\usepackage{graphicx}
\usepackage{amssymb}
\usepackage{amsmath}
\usepackage{bm}
\usepackage{amsmath, amsthm, amssymb}
\usepackage{color}

\def\beq{\begin{equation}}
\def\eeq{\end{equation}}
\def\bea{\begin{eqnarray}}
\def\eea{\end{eqnarray}}

\begin{document}

\title{Designer quantum spin Hall phase transition in molecular graphene}

\author{Pouyan Ghaemi, Sarang Gopalakrishnan, and Taylor L. Hughes}

\affiliation{Department of Physics, University of Illinois at Urbana-Champaign, Urbana, IL 61801, USA}

\begin{abstract}

Graphene was the first material predicted to be a time-reversal-invariant topological insulator; however, the insulating gap is immeasurably small owing to the weakness of spin-orbit interactions in graphene. A recent experiment~\cite{gomes2012} demonstrated that designer honeycomb lattices with graphene-like ``Dirac'' band structures can be engineered by depositing a regular array of carbon monoxide atoms on a metallic substrate. Here, we argue that by growing such designer lattices on metals or semiconductors with strong spin-orbit interactions, one can realize an analog of graphene with strong intrinsic spin-orbit coupling, and hence a highly controllable two-dimensional topological insulator. We estimate the range of substrate parameters for which the topological phase is achievable, and consider the experimental feasibility of some candidate substrates.
\end{abstract}

\maketitle

The seminal work of Kane and Mele predicted the existence of a new state of matter, the quantum spin Hall (QSH) insulator, in monolayer graphene sheets~\cite{kane2005A}. The intrinsic spin-orbit interaction in graphene was shown to generate an energy gap in the nominally semi-metallic band structure, giving rise to an insulating state that is topologically distinct from a conventional band insulator. 
While this prediction spawned the field of time-reversal-invariant topological insulators, it was shown that graphene is a poor candidate material because the insulating bulk gap is extremely small, essentially due to the fact that the carbon atoms, being light elements, have a weak spin-orbit coupling strength~\cite{min,zhang}. A possible way around this was found in Ref. \cite{alicea2011} by depositing adatoms on the graphene surface though it is yet to be experimentally realized. Soon after the Kane-Mele proposal, HgTe/CdTe quantum wells were predicted~\cite{bernevig2006c} and experimentally confirmed~\cite{koenig2007} to exhibit the QSH insulating phase when the thickness of the HgTe layer was tuned properly. Although graphene has a small insulating gap, a realization of the the QSH insulator in graphene would have had certain advantages over the  realization in HgTe/CdTe quantum wells, notably: (i)~being a purely two-dimensional system, graphene is accessible to surface probes, unlike the quantum-well-based systems; (ii)~the phase transition from a trivial to a topological phase in graphene can be tuned via an electric field during the course of an experiment, whereas in the quantum-well systems the tuning parameter is the HgTe layer-thickness, which cannot be tuned in-situ or even continuously. 

With the recent creation of ``molecular graphene'' a new window of opportunity exists to find the QSH insulator in a surface-accessible, electrically tuneable graphene analog~\cite{gomes2012}. Gomes \emph{et al.} showed that when a triangular lattice of repulsive carbon monoxide (CO) molecules is \textit{imposed} on a two-dimensional electron gas (2DEG) (specifically the metallic surface states of copper), the electrons are confined to an artificial honeycomb lattice. 
In fact, the \emph{repulsive} potential generated by a periodic arrangement of the CO molecules always acts to impose the \textit{dual} lattice structure on the underlying electron gas
\footnote{The dual lattice consists of atoms located at the center of every plaquette of the original lattice}. 
In particular, as mentioned, a triangular array of molecules generates a \textit{honeycomb} array of interstices, thus realizing the same lattice symmetries as those of graphene. Remarkably, the experiments of Ref. \cite{gomes2012} showed, via scanning-tunneling microscopy, that the main features of the \emph{electronic} structure of graphene (\emph{e.g.}, the massless Dirac dispersion) are also realized in the ``molecular graphene" analog.

In the present work we argue that the Kane-Mele QSH model can be realized in a molecular graphene system by depositing CO molecules on the surface of a suitable heavy metal or semiconductor with more strongly spin-orbit coupled surface states than copper. We show that the microscopic spin-orbit coupling (SOC) on the surface of the heavy metal generates \textit{both} the competing ``Rashba'' and ``intrinsic'' spin-orbit coupling terms in the low-energy Hamiltonian of Kane and Mele. As the two low-energy terms depend \textit{differently} on the microscopic SOC, their relative sizes (and therefore the nature of the ground state) can be tuned by altering the microscopic SOC via applied electric fields or tuning the CO-generated lattice structure. Thus, the system discussed in the present work realizes the field-tuned transition between the topological insulator and the Dirac metal that was originally predicted by Kane and Mele~\cite{kane2005A}. Furthermore, as we discuss below, the exquisite tunability of molecular graphene makes it possible to study, for example, the physics of interfaces between topological and trivial insulators, in ways that have no parallel in conventional systems.

\begin{figure}
	\centering
		\includegraphics[height=1.2in, width=3.4in]{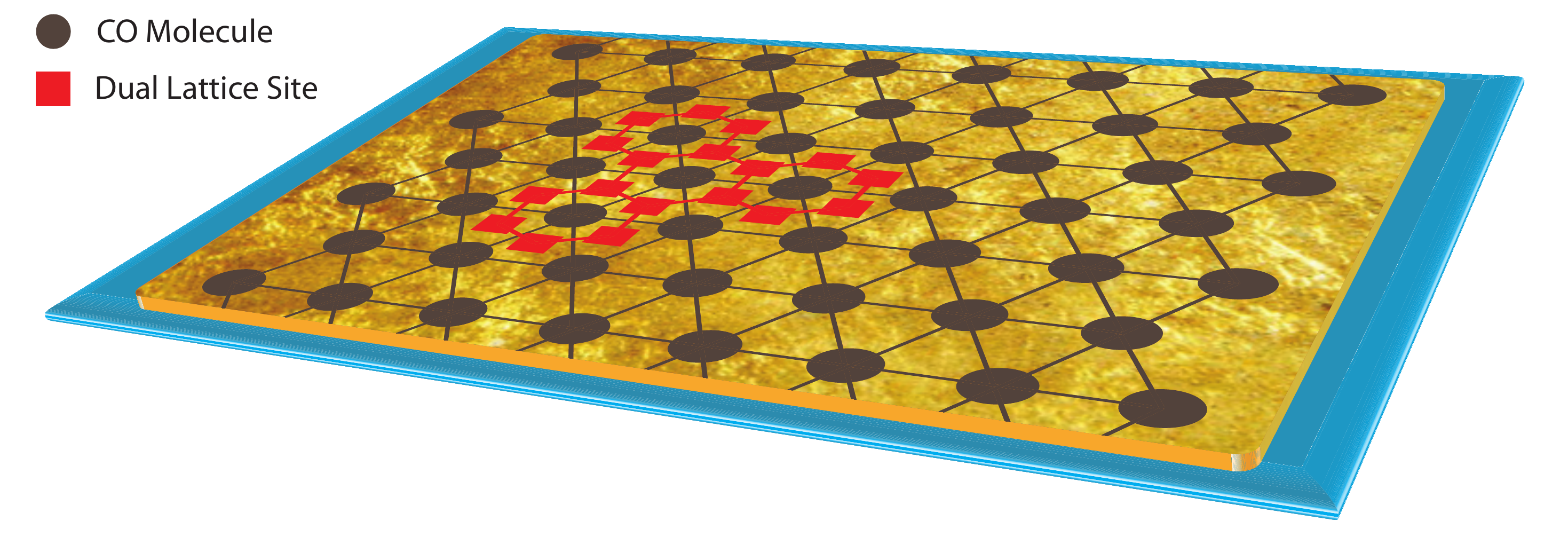}
	\caption{Schematic illustration of proposed experimental setup, involving a triangular array of CO molecules (black circles) placed on a metallic substrate. The surface electrons of the metal are repelled by the CO molecules, and therefore occupy the sites of the dual lattice (red squares), which form a honeycomb lattice.}
	\label{fig:goldsurface}
\end{figure}

We will consider a 2DEG formed by confined states on the (111) surface of a metal or semiconductor such as Ag, Au, or Bi. To be explicit let us consider the Au(111) surface which exhibits surface states which are well-separated in energy from the bulk-states near the $\Gamma$-point of the Brillouin zone (BZ)\cite{kaven}. The surface states arise from  \emph{sp}-hybridized electronic orbitals and are well-described by a simple $s$ and $p$-orbital tight-binding model for the surface orbitals from which, if we ignore SOC, a single, spin-degenerate band is extracted as a low-energy model\cite{petersen2000}. At low energies this degenerate band is the only relevant surface state and can be accurately represented in the isotropic, nearly-free-electron model~\cite{ash} with the Hamiltonian

\begin{equation}
H_{0}=\frac{P^2}{2m^{\ast}}
\end{equation}\noindent where $m^{\ast}$ is an effective mass parameter. 
Following Refs. \cite{park2009,gomes2012}, we consider the addition of a periodic potential 
induced by a triangular lattice of CO molecules placed on the (111) surface (see Fig. \ref{fig:goldsurface}).  We assume that the electrons in the 2DEG can be taken to be noninteracting, so that the microscopic Hamiltonian we consider, $\mathcal{H}$ is of the form
\beq
\mathcal{H} = H_0 +  V_{CO}(\mathbf{r})
\eeq
\noindent where we approximate the molecular periodic potential $V_{CO}(\mathbf{r})$ by its lowest Fourier component,
\beq
V_{CO}(\mathbf{r}) = V_G \sum\nolimits_i \cos(\mathbf{G}_i \cdot \mathbf{r}),
\eeq
where $V_G>0$ is repulsive and the two reciprocal lattice vectors for the triangular lattice are $\textbf{G}_1=\left(\frac{2\pi}{d}\right) \frac{2}{\sqrt{3}} (1/2,-\sqrt{3}/2)$ and $\textbf{G}_2=\left(\frac{2\pi}{d}\right) \frac{2}{\sqrt{3}} (1/2,\sqrt{3}/2)$ for CO molecules separated by a distance $d.$ These reciprocal lattice vectors generate a hexagonal first BZ.

\begin{figure*}
	\centering
		\includegraphics{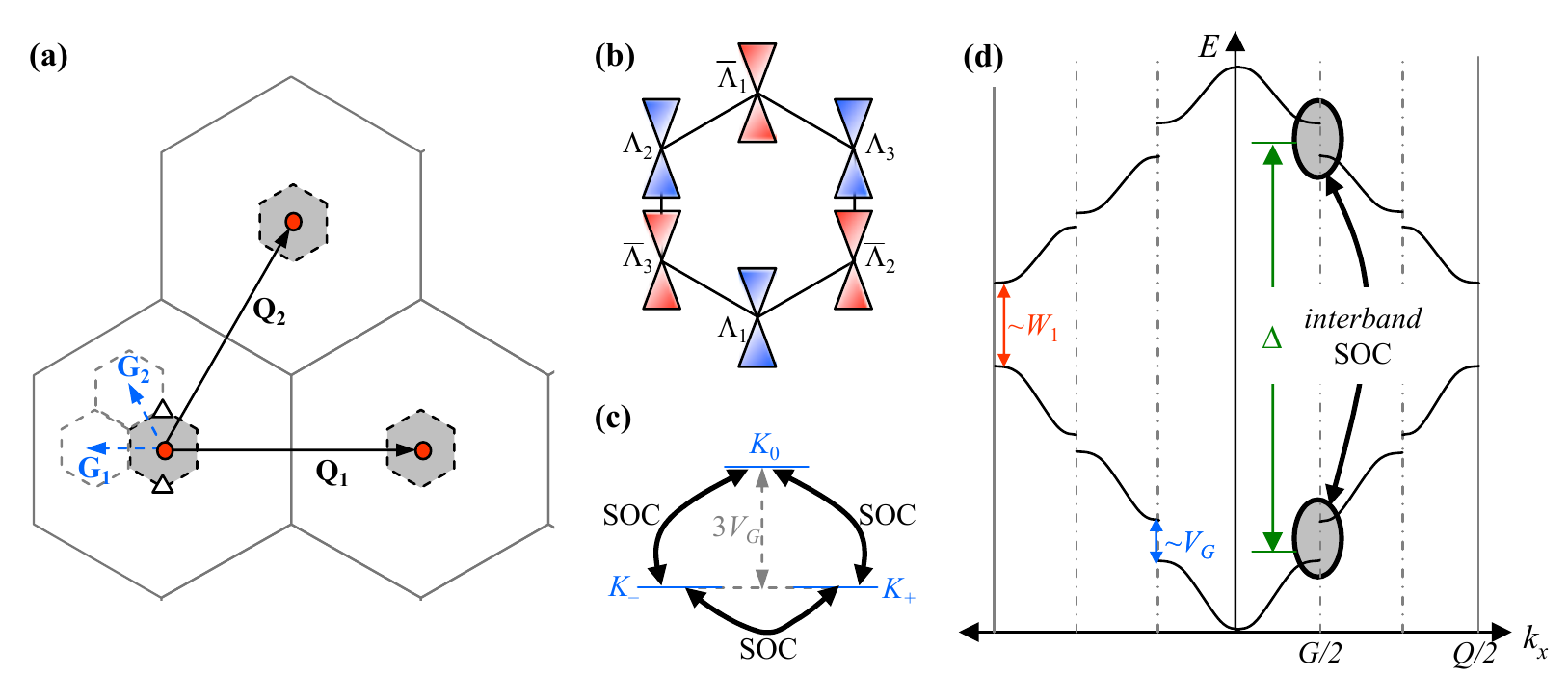}
	\caption{Band structure of ``molecular graphene'' on a periodic (honeycomb-lattice) substrate, e.g., the surface of Au. (a)~Reciprocal lattice structure, showing the substrate reciprocal lattice (gray solid hexagons) as well as the first Brillouin zone (BZ) of the reciprocal ``molecular graphene'' superlattice (dashed shaded hexagons) near the $\Gamma$ point---red dot---of each substrate unit cell. Solid and dashed arrows denote reciprocal lattice vectors of the substrate ($\mathbf{Q}_i$) and of the superlattice ($\mathbf{G}_i$) respectively. The two inequivalent Dirac cones are represented using triangles.  (b)~``Dirac'' band structure for states at corners of the first superlattice BZ. The six corner momenta form two inequivalent groups (blue and red respectively $\Lambda_i,\bar{\Lambda}_i$); (c) states at the three momenta marked (e.g.) in blue hybridized via the lattice potential to form two low-lying states and one higher-energy state. All three states are coupled via the intraband SOC. (d)~Schematic one-dimensional cut through the band structure, showing the band gaps due to the substrate lattice potential ($\sim W_1$) and superlattice potential ($\sim V_G$), as well as the multiple superlattice BZs that fit within the first substrate BZ. We are primarily interested in states at the edges of the first superlattice BZ (shaded ellipse), which are connected by the interband spin-orbit coupling (SOC) to the corresponding momenta in the upper band.}
	\label{fig:bandstruc}
\end{figure*}

The Hamiltonian $\mathcal{H}$ has been carefully analyzed previously\cite{wunsch2008,park2009,gomes2012} and, as we briefly recount now, gives rise to massless Dirac cones. In the nearly-free electron picture the potential $V_{CO}(\mathbf{r})$ couples states with different momenta which are separated by reciprocal lattice vectors. 
To first order in $V_G$, one only needs to consider degenerate free-electron states that are coupled by a single reciprocal lattice vector $\mathbf{G}_i$. Thus, we focus on the neighborhood of the six $k$-points at the corners of the first BZ (see Fig. \ref{fig:bandstruc}a,b).
The six corners naturally decouple into two sets of three ``equivalent''  points connected by reciprocal-lattice vectors denoted $\Lambda_i, \bar{\Lambda}_i$ respectively for $i=1,2,3.$ The original eigenstates are modified by a non-vanishing $V_G$ and the new eigenstates are formed from linear combinations of the Bloch waves at the three $\Lambda_i$ or $\bar{\Lambda}_i.$ We can thus perform degenerate state perturbation theory at these points with the effective Hamiltonian
\begin{equation}\label{ht}
H_{\Lambda}= \left(\begin{array}{ccc} \epsilon(\Lambda_1) & V_G & V_G \\ V_G &\epsilon(\Lambda_2) & V_G\\ V_G & V_G &\epsilon(\Lambda_3)  \end{array}\right) 
\end{equation}
\noindent where $\Lambda_1,\Lambda_2=\Lambda_1-\mathbf{G}_1,\Lambda_3 =\Lambda_1+\mathbf{G}_2$ are one set of the BZ corners and $\epsilon(\Lambda_i)=\frac{\hbar^2}{2m^*} \left(\frac{4\pi}{3 d}\right)^2.$ This Hamiltonian has a $C_3$ symmetry under permutation of the $\Lambda_i$; the eigenstates break up into a singlet  representation of $C_3$ with energy $E_{+}=\epsilon(\Lambda)+2V_G$ and wavefunction  $|K_0 \rangle = \frac{1}{\sqrt{3}}\left(1,1,1 \right),$ and a lower-energy doublet representation with energy $E_{-}=\epsilon(\Lambda)-V_G$ and wavefunctions $|K_{\pm} \rangle = \frac{1}{\sqrt{3}}\left(1, e^{\pm 2i\pi/3},e^{\pm 4i\pi/3}\right).$
As shown in Refs. \onlinecite{wunsch2008,park2009,gomes2012}, if one projects the Hamiltonian onto the low-energy subspace spanned by $| K_{\pm} \rangle$ and considers the $k\cdot P$ Hamiltonian perturbed around the $\Lambda_i$ then one arrives at the massless Dirac Hamiltonian. If we combine the doublet $|K_{\pm}\rangle$ from the $\Lambda_{i}$ points  with the doublet $|K'_{\pm}\rangle$ from the $\bar{\Lambda}_i$ points  we arrive at the effective graphene-like Hamiltonian 
\begin{equation}\label{gr}
H_{\mathrm{Dirac}}=\hbar v_f \left[k_x \sigma_1+k_y \tau_3\sigma_2\right]
\end{equation}
where $v_f=\frac{1}{3m^{\ast}}\frac{2\pi\hbar}{d},$ $\tau_3$ is a $2 \times 2$ Pauli matrix acting on the valley index $(K,K'),$ $\sigma_a$ are the Pauli matrices acting on the doublet $\pm$ index, and the tensor product is implicit. 

 In addition to the effective projected model stemming from $H_0,$ we wish to consider the effects of spin-orbit coupling on the low-energy electronic structure. 
 Without SOC (and without external magnetic fields) we find that $H_{\mathrm{Dirac}}$ is trivially extended to a spin-independent $8\times 8$ Hamiltonian which includes the spin-$1/2$ degree of freedom. Electric fields generated by surface effects and the atomic crystalline potential will give rise to non-trivial spin-orbit effects in $H_{\mathrm{Dirac}}.$ Angle-resolved photo-emission experiments clearly show a large spin-splitting in the Au(111) surface states due to a non-vanishing Rashba effect described by the free-electron Hamiltonian of the  surface states with SOC
 \begin{equation}\label{Hrashba}
 H^{(SOC)}_0=\frac{P^2}{2m^{\ast}}+\alpha_R \left({\textbf{P}}\times{\textbf{s}}\right)\cdot \hat{z}
 \end{equation}\noindent where $\alpha_R$ is the Rashba coefficient and ${\textbf{s}}$ is a set of Pauli matrices representing physical spin-1/2\cite{LaShell}. Projecting this term onto the low-energy Dirac subspace involves certain subtleties, which arise 
because (as is well known) the effective SOC in a projected band arises from transitions to neighboring bands\cite{winklerbook}. Thus, since the symmetries of Eqs. \ref{gr} and \ref{Hrashba} are quite different, the form of the terms that can appear in Eq. \ref{gr} will potentially also be different. 
 
 Thus, with these caveats in mind, we will derive the effective spin-orbit coupling seen by the low-energy Dirac Hamiltonian from ``first principles'' in the nearly-free electron limit. From symmetry considerations we expect two types of spin-orbit terms will arise in Eq. \ref{gr}: (i) a Rashba-type term arising from the inversion symmetry breaking on the Au substrate surface (ii) an intrinsic SOC term of the Kane-Mele type. As we will now show, in the nearly-free electron limit both of these terms appear naturally if we include the effects of the surface potential 
 \beq\label{vsubs}
V_{subs.}(\mathbf{r}) =  \Theta(-z) \left[W_0 + W_1 \sum\nolimits_i \cos(\mathbf{Q}_i \cdot \mathbf{r})\right]
\eeq
 where $\Theta(-z)$ is a step function (since we have chosen the Au to be in the region $z<0$), and the periodic surface potential of the Au(111) crystal substrate is taken to have a hexagonal surface BZ with the reciprocal lattice vectors $\textbf{Q}_1=\left(\frac{2\pi}{a}\right) \frac{2}{\sqrt{3}} (1/2,-\sqrt{3}/2)$ and $\textbf{Q}_2=\left(\frac{2\pi}{a}\right) \frac{2}{\sqrt{3}} (1/2,\sqrt{3}/2)$ where $a$ is the lattice constant of $Au.$ The $W_0$ term essentially represents the work-function due to charge accumulation near the Au(111)/vacuum interface and the $W_1$ term is the periodic modulation induced by the crystalline potential of the Au atoms.  A few notes are in order at this point: (i) instead of including the $W_1$ term we could have reformulated our electronic structure calculation using the $sp$-orbital tight-binding model of Ref. \onlinecite{petersen2000} which would naturally generate the higher surface-bands needed to properly account for the SOC, but we chose to stay with the nearly-free electron formalism,  where higher bands are generated by the Au lattice potential ($\propto W_1$), for a unified presentation; (ii) we have assumed a hexagonal surface BZ geometry  as in an Au(111) crystal-surface\cite{Mazarello}; this assumption simplifies our analysis but is not an essential feature, as it will not qualitatively affect our analysis if the BZ geometry is different. We can see that the latter is true by noting that since $a\ll d$ then $\vert Q_i\vert\gg \vert G_i\vert$ so the effective Dirac structure lies well within the first BZ of the Au(111) periodic potential and thus the details of the substrate bandstructure at the lattice scale cannot substantially affect $H_{\mathrm{Dirac}}$ (see Fig. \ref{fig:bandstruc}a).

For free electrons the general microscopic expression for the SOC is
\beq
H_{SOC} = \frac{\alpha_0}{\hbar} (\nabla V_{subs.} \times \mathbf{P}) \cdot \mathbf{s}
\eeq
where $\alpha_0$ is the bare spin-orbit coupling, $\mathbf{P}$ denotes the momentum, and $\mathbf{s}$ the physical spin. Using the form of $V_{subs.}$ in Eq.~(\ref{vsubs}), we can split the contributions into nominally Rashba, and intrinsic pieces
\begin{eqnarray}
H_R &=& \frac{\alpha_0}{\hbar \xi} W_0 (\mathbf{\hat{z}} \times \mathbf{P})  \cdot  \mathbf{s}\\
H_{I}& =& \frac{\alpha_0}{\hbar \xi} W_1   (\mathbf{\hat{z}} \times \mathbf{P}) \cdot \mathbf{s} \, \sum\nolimits_i \cos(\mathbf{Q}_i \cdot \mathbf{r}) + \ldots  
\end{eqnarray}
where $\xi$ is the localization length of the surface states along $z$, and the omitted terms in $H_I$ involve in-plane gradients of the surface periodic atomic potential which do not generate any nontrivial SOC terms within the nearly-free-electron scheme
. Owing to the form of its spatial modulation, which depends on the large reciprocal lattice vectors ${\textbf{Q}}_i(\sim 1/\AA),$ $H_{I}$ only has non-vanishing matrix elements between states in (what we call) different  \emph{substrate} bands. It is analogous to the microscopic SOC considered in the tight-binding model of graphene in Ref.~\cite{min} which takes into account transitions to other high-energy bands.

To project these SOC terms onto the subspace of the Dirac Hamiltonian we will use perturbation theory in the strength of the SOC. To carry this out we must assume that the SOC energy scale is smaller than the energy scale that gives rise to the gapless Dirac structure set by $V_G$, and which in turn is assumed smaller than the free-electron energy scale set by $\hbar^2 |G_{1}|^2/2m^{*}.$ Thus, we treat the lattice as a perturbation acting on the free-electron states to generate the Dirac subspace, and then treat the SOC as a perturbation acting on the gapless Dirac Hamiltonian. The assumption that the SOC is weaker than the other energy scales is essential to our analysis; however, the assumption of a weak lattice can presumably be relaxed without changing our qualitative conclusions. 

We will proceed in two steps by projecting $H_R$ first, and then $H_I.$ Projecting $H_R$ (which includes the dominant contributions from a Lowdin partitioning of the higher bands\cite{winklerbook} ) we find contributions to an effective Rashba coupling \emph{and} an effective Kane-Mele coupling in $H_{\mathrm{Dirac}}$ of the form
\begin{eqnarray}
H^{R}_{eff}&=&\lambda_{RR}\left(\tau_3\sigma_2 s_1+\sigma_1 s_2\right)\\
H^{I}_{eff}&=&\lambda_{IR}\tau_3\sigma_3 s_3
\end{eqnarray}\noindent where $\lambda_{RR}=(1/2)W_0\gamma,$ $\lambda_{IR}= \frac{\gamma^2 W_0^2}{6 V_G },$ and $\gamma=\alpha_0\vert\Lambda_1\vert/\xi$  where we have included the leading intra-substrate band contributions, dropped an unimportant constant in $H^{I}_{eff},$ and used the fact that the surface state wavefunctions exponentially decay into the bulk with a characteristic length $\xi$  (see Fig. \ref{fig:bandstruc}c,d for the origin of the intra-substrate band contributions). The terms generated by $H_R$ within the low-energy theory are thus of two kinds: (i)~a Rashba-term that is first-order in $W_0$ and arises because $H_R$ mixes the two low-lying states, and (ii)~ a Kane-Mele term that is second-order in $W_0$, and arises because of virtual transitions into higher bands (but still in the lowest substrate band) mediated by $H_R.$ 
Note that when $\gamma W_0 < 3 V_G$, which is the condition for validity of our perturbation treatment, $\lambda_{IR}< \lambda_{RR}$; therefore, $H_R$ on its own does not give rise to the QSH phase. 
Having addressed the terms generated by $H_R$, we now turn to the spatially modulated coupling $H_{I}$. As we have noted above, due to the spatial modulations in $H_I$ only states between different substrate bands are connected (see Fig. \ref{fig:bandstruc}d). Again, after projecting, we find contributions to the Rashba and Kane-Mele terms with coefficients $\lambda_{RI}=2W_{1}^2\gamma/\Delta$ and $\lambda_{II}=2 W_{1}^2\gamma^{ 2}/\Delta$ respectively where $\Delta=\hbar^2 \vert Q\vert^2/2m^{\ast}$.

\begin{figure}
	\centering
		\includegraphics[height=2.5in, width=3.4in]{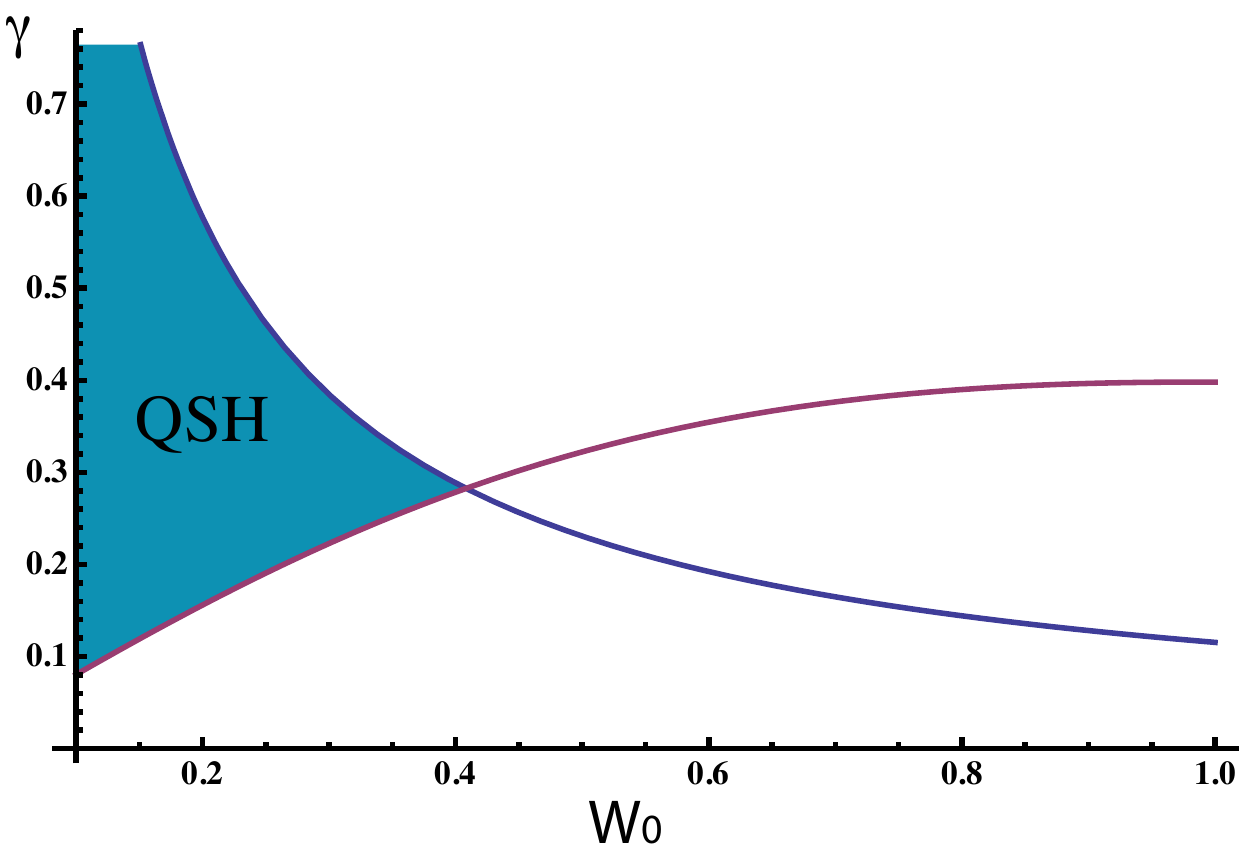}
	\caption{Phase diagram of ``molecular graphene'' as a function of the spin-orbit coupling (SOC) parameter $\gamma$, introduced in the main text, and the work function $W_0$ of the substrate (which can be tuned to some extent via gating). The quantum spin Hall insulator is achieved in the region shaded blue.}
	\label{fig:gammap}
\end{figure}

We can combine these two results to obtain the total effective Rashba and Kane-Mele couplings $\lambda_{R}=\lambda_{RR}+\lambda_{RI}$ and $\lambda_{I}=\lambda_{IR}+\lambda_{II}.$ To observe a topological insulator phase we need $\lambda_{I}>\lambda_{R}$  (along with $\gamma <\frac{3V_G}{W_0+4W_0 W_1/\Delta}$ in order to stay within the effective low energy sector) which imposes the restriction
\begin{eqnarray}
\frac{1/2+2 W_1 /\Delta}{W_0/6V_G+2 W_1^2/W_0 \Delta}<\gamma<\frac{3 V_G}{W_0+4 W_1 W_0/\Delta}.\label{eq:ineq}
\end{eqnarray}
\noindent  The inequality in Eq. \ref{eq:ineq}, which indicates that the dimensionless parameter $\gamma$ is the crucial control parameter, is the primary result of this work. There are three ways to optimize $\gamma=\alpha_0\vert\Lambda_1\vert/\xi$ due to its parameter dependence: (i) use heavy atom materials with large atomic SOC $\alpha_0$; (ii) tune the lattice constant of the CO molecules to be shorter so that $\vert \Lambda_1\vert$ increases; (iii) add surface dopants or electrostatic gating to optimize the effects of the surface potential in the z-direction on the surface state wavefunctions. For our choice of a step function potential this implies we should decrease $\xi$ so that more weight lies near the region of strong electric field ($z=0$). In Fig. \ref{fig:gammap} we plot an approximate phase diagram as a function of $W_0$ and $\gamma$ using the values $\Delta= 5 \mathrm{eV}, W_1= 2 \mathrm{eV}, V_G= 0.1 \mathrm{eV}.$ The phase diagram is shaded where Eq~(\ref{eq:ineq}) holds and the decreasing (increasing) curve represents the right (left) side of the inequality. Note that, for these parameters, the work function $W_0$ must be less than $0.4$ eV in order for the QSH phase to be realizable; this value is much smaller than the work function of pure metals such as Au ($W_0 \sim 5$ eV), but can presumably be tuned, especially for a semiconductor substrate, either chemically (see, e.g., Ref.~\cite{broker}) or via a strong perpendicular electric field. 
For an Au(111) surface, and a CO molecule spacing of $d \approx 3 \AA$, we estimate $\gamma$ of order $.02$ which is probably too small to be in the QSH phase for realistic parameters. On the other hand, there are multiple other compounds (\emph{e.g.} $BiAg_2$ \cite{Frantzeskakis}, $BiTeI$ \cite{ishizaka}, or  $Bi_x Pb_{1 - x} /Ag$ \cite{Christian}) with similar surface band structure, and with a $\gamma >0.2$ for a $d \sim 3\AA$; these should be possible to tune to the QSH phase. One would prefer to have a system near the lower phase boundary so that the topological phase transition could be tuned using an electric gate to nominally adjust $W_0$ or the surface-state parameter $\xi$ entering $\gamma$.

We now turn to possible schemes for detecting and manipulating the topological phase in molecular graphene. The detection method used in the experiments of Ref.~\cite{gomes2012} was scanning-tunneling microscopy (STM), which directly probes the local density of states as a function of energy; STM can reveal many of the important features of the Kane-Mele model as well, such as the existence of a bulk gap and of topologically protected edge states. One can test whether the edge states are protected against backscattering, e.g., by placing ``impurities'' such as vacancies on the edge (this is straightforward to do as the lattice is engineered in the first place) and using STM, either to detect whether the impurities gap out the edge states, or to measure the scattering matrix via the techniques in Ref.~\cite{roushan2009}. Additionally, the ease of engineering sharp interfaces between the topological insulator and a \textit{trivial} insulating phase by increasing the CO spacing in a neighboring region will allow for a careful study of the topological interface states.

To summarize, we have argued that molecular graphene grown on the surfaces of heavy metals or semiconductors should realize the Kane-Mele model, owing to the strong spin-orbit coupling in the underlying metal. The topological phase of the Kane-Mele model can be realized in a parameter range that is experimentally viable and which can be optimized by tuning the lattice spacing, electrical gates, and atomic spin orbit coupling. The tunability and controllability offered by molecular graphene, relative to the existing (quantum-well-based) 2D topological insulators, make it a particularly appealing platform for studying various questions of practical or conceptual interest especially since it provides an open surface with which one can easily access the edge-state properties, and optimistically the ability to tune the QSH topological phase transition by gating. 

We would like to thank P. Goldbart, N. Mason, W. K. Park, and D. Van Harlingen for useful discussions. PG is supported by the ICMT at UIUC. SG and TLH are supported by  the DOE under grant DE-FG02-07ER46453.

\end{document}